\documentclass[9pt, conference]{IEEEtran}

\IEEEoverridecommandlockouts
\usepackage{cite}
\usepackage{amsmath,amssymb,amsfonts}
\usepackage{algorithmic}
\usepackage{graphicx}
\usepackage{textcomp}
\usepackage{xcolor}
\usepackage{bm}
\def\BibTeX{{\rm B\kern-.05em{\sc i\kern-.025em b}\kern-.08em
    T\kern-.1667em\lower.7ex\hbox{E}\kern-.125emX}}

\usepackage[T1]{fontenc}
\usepackage[utf8]{inputenc}
    
\newcommand{\eat}[1]{}

\newcommand{\comment}[1]{\textbf{\color{red}}}
\input{Qcircuit}
\usepackage[flushleft]{threeparttable}
\begin{document}

\title{Ever more optimized simulations of fermionic systems\\  on a quantum computer
\thanks{${}^*$The first two authors contributed equally to this work.}
}

% \eat{
\makeatletter
\newcommand{\linebreakand}{%
  \end{@IEEEauthorhalign}
  \hfill\mbox{}\par
  \mbox{}\hfill\begin{@IEEEauthorhalign}
}
\makeatother

% \eat{
\author{
\IEEEauthorblockN{1\textsuperscript{st} Qingfeng Wang$^*$}
\IEEEauthorblockA{\textit{Joint  Quantum  Institute} \\
\textit{University of Maryland}\\
College Park,Maryland 20742,  USA \\
keewang@umd.edu}
\and
\IEEEauthorblockN{2\textsuperscript{nd} Ze-Pei Cian$^*$}
\IEEEauthorblockA{\textit{Joint  Quantum  Institute} \\
\textit{University  of  Maryland}\\
College  Park,  Maryland  20742,  USA \\
zpcian@umd.edu}
\and
\IEEEauthorblockN{3\textsuperscript{rd} Ming Li}
\IEEEauthorblockA{\textit{{Atom Computing, Inc.}} \\
Berkeley, California 94710, USA \\
lllraistlin@gmail.com}
\linebreakand
\and
\IEEEauthorblockN{4\textsuperscript{th} Igor L. Markov}
\IEEEauthorblockA{\textit{Nova Ukraine} \\
Mountain View, CA, 94040 USA \\
igor.markov@novaukraine.org}
\and
\IEEEauthorblockN{\, }
\IEEEauthorblockA{\textit{{\, }} \\
}
\and
\IEEEauthorblockN{5\textsuperscript{th} Yunseong Nam}
\IEEEauthorblockA{\textit{Department  of  Physics} \\
\textit{University  of  Maryland,  College  Park}\\
Maryland  20742,  USA\\
ynam@umd.edu}
}
% }

\maketitle

\begin{abstract}
Despite using a novel model of computation, quantum computers break down programs into elementary gates. Among such gates, entangling gates are the most expensive. In the context of fermionic simulations, we develop a suite of compilation and optimization techniques that massively reduce the entangling-gate counts. We exploit the well-studied non-quantum optimization algorithms to achieve up to 24\% savings over the state of the art for several small-molecule simulations, with no loss of accuracy or hidden costs. Our methodologies straightforwardly generalize to wider classes of near-term simulations of the ground state of a fermionic system or real-time simulations probing dynamical properties of a fermionic system. 
\end{abstract}

\section{Introduction}
Simulating microscopic phenomena is anticipated to be the epitome of quantum-computing applications. Chemistry, condensed matter, high energy physics and other fields stand to benefit from accurate quantum simulations of {\it fermions}, particles ubiquitous in nature.
To make this endeavor practical, quantum programs simulating fermionic matter must be optimized, or else quantum noise and decoherence may overwhelm quantum calculations. Due to scarce computational resources of a near-term quantum computer~\cite{nisq}, variational quantum eigensolver (VQE)~\cite{VQE} is widely used to estimate the {\em ground state energy} of a molecule, critical for determining its chemical properties, using relatively short programs.
The objective here is to enable the estimation using the least possible amount of quantum computational resources.
Specifically, the number of two-qubit CNOT gates used is reduced. Among the de-facto quantum gateset of CNOT{+}single-qubit gates, effectuating high-quality CNOT gates is the most challenging for physical implementation, both time- and fidelity-wise. Indeed, not only is the computational efficiency improved, is the fidelity of quantum computation also boosted, as the CNOT count is reduced.

{\it Prior art --}
References~\cite{Automated,Cong} focus on the quantum circuit optimization with qubit-to-qubit connectivity and hardware constraints in mind. Reference~\cite{TrotterOrder} leverages execution sequences of quantum subroutines to minimize algorithmic errors incurred.
References~\cite{Marco,H10}
exploit various symmetries of chemistry problems to better optimize quantum circuits. 
Most closely related to our work are \cite{ynam_NPJ_water,hmp2}, where a complete-graph connectivity between qubits, available in trapped-ion~\cite{11QBenchmark} or neutral-atom~\cite{ramette2021any} quantum computers, is considered. Explored and exploited include quantum subroutine execution sequence degrees of freedom and quantum data compression, among others, to significantly optimize chemistry simulations on a near-term quantum machine.

%To this end,
We make the following contributions:
\begin{itemize}
\item For several state-of-the-art simulation methods where circuits are designed hierarchically, 
we streamline and reduce the hierarchy with high-level optimization methods, to reduce gate counts.
\item We perform high-level circuit optimizations using advanced computational tasks (such as the Travelling Salesman Problem and Graph Coloring)
  that we solve with high-performance solvers.
\item For simulations of near-term interest, we reduce published gate counts by 3.5-24\%.
\item We demonstrate optimizations applicable to wider classes of near-term simulations and real-time dynamics simulations.
\end{itemize}

\section{The Baseline VQE algorithm}
\label{sec:Prelim}

\begin{figure}
\includegraphics[scale=0.2]{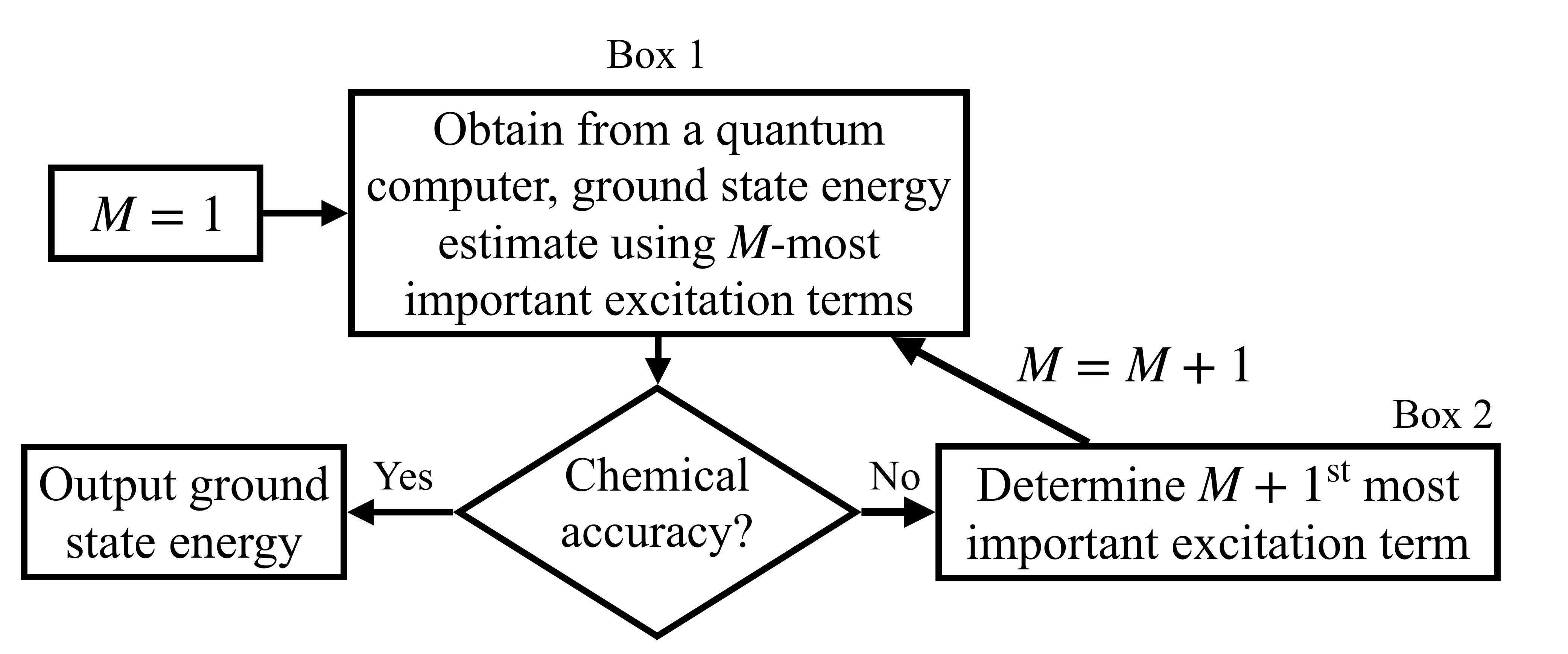}
\caption{\label{fig:VQE} Typical VQE cycle. $M$ is a parameter that denotes the size of the ansatz, such as the number of excitation terms. The expressivity of the ansatz is increased as the size becomes larger. Quantum computers here provide the ground-state energy estimate, given the ansatz specification. The ansatz size is gradually increased until it is sufficiently large to allow for the energy estimate to be within a pre-specified threshold from the ground sate energy, such as chemical accuracy. Other termination criteria, including requiring a minimum energy-estimate improvement, can be considered as well.}
\end{figure}

Fig.~\ref{fig:VQE} illustrates the computational loop of the Variational Quantum Eigensolver (VQE) algorithm~\cite{VQE}, which constructively estimates the min-value of a function (energy), with real-valued input variables, evaluated on a quantum computer. To start, choose an {\it excitation} term, kicking the electrons around in a molecule. This explores the landscape of the function in one direction by setting the corresponding {\it variational parameter} to nonzero, to set how far in the chosen direction is the exploration done. This prepares a parametrized ``guessed'' ground state on a quantum computer. The prepared state can be used to determine its energy. The energy obtained can inform the search for an optimal parameter that admits the lowest energy. The lowest energy found can be used to decide if an additional excitation term (direction) needs to be considered to be sufficiently expressive in approximating the ground state. The cycle stops once the energy found is within an acceptable threshold (e.g., chemical accuracy, an expected energy fluctuation due to room-temperature environment) to the ground-state energy.

{\it Technical detail} -- VQE outputs the ground-state energy estimate $E {=} \langle \Psi | H |\Psi \rangle $, where $H$ is the system Hamiltonian (energy operator). The variational ansatz state $|\Psi \rangle $ is constructed from a product state $| \Psi_0 \rangle $ using a unitary transformation induced on a quantum computer, $|\Psi \rangle {=} U |\Psi_0\rangle $, where $U {=} \exp(Z {-} Z^\dagger )$. The fermionic excitation operator $Z$ contains the variational parameters $\theta_{\vec{\gamma},\vec{\alpha}}$ and is of the form
$
    Z {=} \sum \theta_{\vec{\gamma},\vec{\alpha}} \bigotimes_{\vec{\gamma}} c_\gamma^\dagger
    \bigotimes_{\vec{\alpha}} c_\alpha,
$
where $\vec{\gamma}$ and $\vec{\alpha}$ denote sites of a molecule or orbitals in which electrons are to be created ($c_\gamma^\dagger$) and annihilated ($c_\alpha$). By varying the parameters $\theta$ in $Z$, $E$ is variationally minimized.
Due to its versatility, a {\it unitary coupled cluster single double} (UCCSD) ansatz $U$ is often used, where $Z{=} Z_1 {+} Z_2$,
\[
    Z_1 =\sum_{ p \in {\rm V},~r \in {\rm O}} \theta_{pr} c^\dagger_p c_r, \quad
    Z_2 = \sum_{p,q \in {\rm V},~r,s \in {\rm O}} \theta_{pqrs} c^\dagger_p c^\dagger_q c_r c_s. 
\]
``V'' and ``O'' here denote virtual (empty) and occupied (filled) molecular spin orbitals. To be complete, $|\Psi_0\rangle$ can be a ground-state Hartree-Fock wavefunction, multiconfigurational self-consistent field~\cite{mcscf}, or density-matrix renormalization group calculations~\cite{dmrg}. Many choose to work with the former for its simplicity, including \cite{ynam_NPJ_water,hmp2} and this work.

{\bf Baseline optimization} --
Focusing on Box~1 of Fig.~\ref{fig:VQE}, \cite{ynam_NPJ_water,hmp2} investigated the compilation and optimization of the UCCSD ansatz, where, by use of the first-order {\it Trotter formula}~\cite{ar:Suzuki}, the circuit implemented was of the form 
$\prod_{k} e^{O_k},$
where each $O_k$ denotes individual summands of $Z{-}Z^\dagger$.
At a high level, (overly-)complicated optimization techniques used in \cite{ynam_NPJ_water,hmp2} can be summarized as:

{\it ``Bosonic'' encoding} -- 
Exploiting quantum data compression, when the state of an electron pair is symmetric, e.g., a superposition of $|00\rangle$ and $|11\rangle$, one can use only a single qubit, not two, to encode the state of the pair on a quantum computer. This also compresses the quantum operations that involve the pair. Previously, the compression was used only when two such symmetric pairs were provided simultaneously\footnote{The compressed operations follow the so-called {\em hard-core bosonic algebra}, hence the name ``Bosonic'' encoding.}.

{\it Fermion-to-qubit transformation matrix} -- Operators that create and annihilate fermions obey certain rules. On the other hand, operators that excites ($\ket{0}{\rightarrow}\ket{1}$) and de-excites ($\ket{1}{\rightarrow}\ket{0}$) a qubit obey different rules. Therefore, fermionic operators need to be properly transformed to qubit operators (e.g., Pauli matrices $X,Y,Z$), prior to their implementation on a quantum computer. Here, the choice of transformation drastically changes the computational resource requirement. Previously, the space of an $n {\times} n$ {\it upper-triangular} reversible binary matrix, which can represent a subset of valid transformations, was searched using particle swarm optimization.

{\it Fermionic level labeling} -- The embedding of electronic sites onto qubits is another choice that can drastically change the resource requirement. Due to factorially large embedding space, a simple greedy approach that tries a transposition at a time was used.

{\it Intra-excitation term ordering} -- Each term implemented in a quantum circuit was $e^{O_k}$ (see above). $O_k$, once transformed to qubit operators, becomes a sum of Pauli strings, a tensor product of Pauli matrices. $e^{O_k}$ was implemented by applying the circuit that implements the matrix exponentiation of each Pauli-string summand, one after another. An exhaustive search was performed to find the best ordering of the summands for each $e^{O_k}$.

{\it Target qubit choice} -- Notice both circuits implement $e^{i\theta ZZ/2}$:
\[
  \Qcircuit @C=0.5em @R=.1em @!R {
&\empty &\qw &\ctrl{1} &\qw                &\ctrl{1} &\qw &\qw      
\\
&\empty &\qw &\targ    &\gate{R_z(\theta)} &\targ    &\qw &\qw    
}
\hspace{3em}
  \Qcircuit @C=0.5em @R=.1em @!R {
&\empty &\qw &\targ &\gate{R_z(\theta)} &\targ &\qw &\qw      
\\
&\empty &\qw &\ctrl{-1} &\qw &\ctrl{-1} &\qw &\qw    
}
\vspace{-0.5em}
\]
The target qubit choice ($\oplus$ of CNOT) in the circuit-level implementation can in fact result in different gate cancellations. Previously, all Pauli strings from the same $O_k$ term shared the same target.

{\it Inter-excitation term ordering} -- The relative ordering between $O_k$ can also be used to optimize quantum circuits, aiming to expose as much similarity between neighboring $O_k$'s. Previously, a doubly-greedy approach was used, one to group as many $O_k$'s implementable with the same target and the other to find the best $O_k$-ordering within each group, to find locally-optimized circuits.

 For Box~2 of Fig.~\ref{fig:VQE},
 recall in a typical VQE approach, one simply measures the expectation values of a set of operators that correspond to the Hamiltonian. Shown in \cite{hmp2} was that this is nothing but first-order perturbation theory. Therefore, by additionally measuring expectation values of yet another set of operators that correspond to second-order corrections, the energy estimate is improved, and perhaps more importantly, which excitation term would need to be additionally considered to improve ground-state energy estimate is determined. Indeed, the rapid convergence of the energy estimate afforded by it in turn resulted in needing a fewer number of excitation terms to reach a pre-specified convergence goal, such as the oft-discussed chemical accuracy in the literature~\cite{chemical_accuracy}.

 {\it Remarks} -- The baseline approach for Box 1 of Fig.~\ref{fig:VQE} detailed in this section will be contrasted with our proposed methods in the next section.
 For Box 2 of Fig.~\ref{fig:VQE}, our work also employs the second-order perturbation theory based approach (i.e. HMP2 in \cite{hmp2}).

\section{Proposed Methodologies}
\label{sec:Method}

\begin{figure}[t]
\includegraphics[width=9cm]{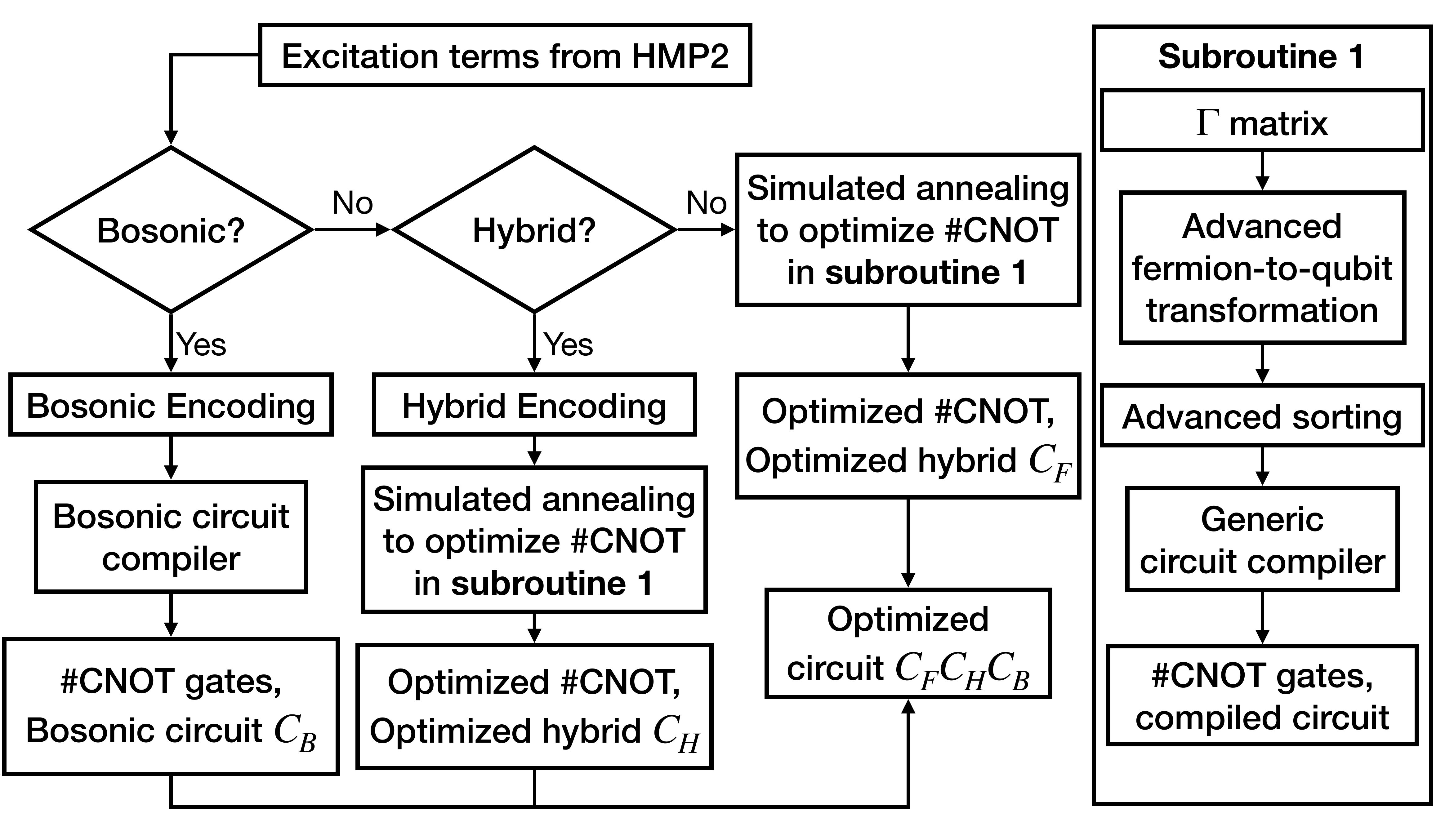}
\caption{\label{fig:evermore} Our procedure to compile and optimize a VQE circuit. }
\centering
\end{figure}

We drastically improve the quality of optimization while massively streamlining the procedure. In contrast to the baseline, we use:
\begin{enumerate}
\item {\it Hybrid encoding}: a generalization of bosonic encoding to encompass partial quantum information compressibility for arbitrarily many orbitals, 

\item {\it Advanced sorting}: a proper superset of the union of Intra-excitation term ordering, Target qubit choice, and Inter-excitation term ordering, and

\item {\it Advanced fermion-to-qubit transformation}: a proper superset of the union of fermion-to-qubit transformation matrix and fermionic level labeling.
\end{enumerate}
Further, our methodologies are based on well-studied graph properties and optimization problems, obviating the need for custom heuristics, amenable to the use of powerful commercial solvers.

Fig.~\ref{fig:evermore} shows a complete flow diagram for our advanced compilation and optimization methodology for VQE simulations of fermionic systems. Briefly, for a given set of excitation terms, we first classify the excitation terms into bosonic, hybrid, and fermionic classes (see Sec.~\ref{sec_hybrid_encoding}), depending on the parity symmetry of each excitation terms. 
The bosonic terms can be simply compiled to quantum circuits using the known procedure described in  \cite{ynam_NPJ_water}. For the hybrid and fermionic terms, we use the procedures described in Secs.~\ref{sec_hybrid_encoding} and \ref{sec_fermion_GTSP}. Recall, in order to compile the unitary operators into a quantum circuit, a choice of fermion-to-qubit transformation must be made. We search for the transformation that admits the best optimization using the procedure described in Sec.~\ref{sec_search_gamma}. 
We refer the readers to the appendix for concrete examples.

\subsection{Hybrid Encoding}
\label{sec_hybrid_encoding}
Consider a state that is a linear superposition of $\ket{00}$ and $\ket{11}$, or $\ket{01}$ and $\ket{10}$. There is exactly one quantum bit of information encoded, so a single qubit is sufficient. 
Now consider a system of $n$ qubits where, by the circuit elements to be applied and with the knowledge of the input state, a two-qubit subsystem state can be predicted to be a superposition of $\ket{00}$ and $\ket{11}$, or $\ket{01}$ and $\ket{10}$. Instead of two qubits, using only one qubit suffices. The wave function satisfying this property have number parity symmetry. Specifically,
$P_{ij}|\psi\rangle = \pm |\psi\rangle,$
where $P_{ij} = (-1)^{c^\dagger_ic_i + c^\dagger_jc_j}$ is the number parity operator for spin orbitals $i$ and $j$.

Any symmetry-preserving operator over two qubits that commutes with $P_{ij}$ can be compressed to a single-qubit operator. The compressed encoding can be recovered to a two-qubit state through a CNOT gate (up to single qubit gates), i.e.,  $CNOT(a\ket{0}+b\ket{1})\ket{0} \mapsto a\ket{00}+b\ket{11}$ for an even-parity compression and  $X_{1}CNOT(a\ket{0}+b\ket{1})\ket{0} \mapsto a\ket{10}+b\ket{01}$ for an odd-parity compression.

Applied to two pairs of subsystems that each contains the symmetry, explored in \cite{ynam_NPJ_water} was the so-called bosonic encoding. Consider a double-excitation unitary $\exp(\theta c^\dagger_p c^\dagger_{q} c_r c_{s} -h.c.)$ on $|\psi\rangle$, an eigenstate of the number parity operators $P_{pq}$ and $P_{rs}$ with eigenvalues $+1$.
Leveraging the parity symmetry and defining the so-called hard-core bosonic operators $d_{pq}^\dagger := c_p^\dagger c_{q}^\dagger$ and $d_{rs} := c_r c_{s}$, $d^\dagger_{pq}$ and $d_{rs}$ map to the Pauli raising/lowering operators $\sigma^{\pm}$. The reduced, ``bosonic'' unitary operator is then $\exp(\theta \sigma_{p}^+ \sigma_{r}^- -h.c.)$.

Consider a different $|\psi\rangle$, an eigenstate of the number parity operator $P_{pq}$ with eigenvalue +1, but not an eigenstate of the operator $P_{rs}$. Following the procedure above, we can reduce the excitation operator to $\sigma^+_p c_r c_{s}$, a one-qubit and two-fermion operator. The operator is hereafter referred to as a hybrid double excitation term.

Note a hybrid double excitation term, in order for it to be reduced, requires the input state to contain the symmetry. This raises an important challenge: Ordering of the excitation terms in the ansatz circuit. For example, consider two double excitation terms $h_1 = c^\dagger_2 c^\dagger_3 c_5 c_{6}$ and $h_2 = c^\dagger_4 c^\dagger_5 c_7 c_8$ and consider an input state $\ket{\psi}$ with the pair symmetry on (5,6), i.e., $P_{56}|\psi\rangle = |\psi\rangle$. Applying $e^{\theta_1 h_1 - h.c.}$ first, the (5,6) symmetry is preserved, hence compressible. Apply $e^{\theta_2 h_2 - h.c.}$ first, then try to apply $e^{\theta_1 h_1 - h.c.}$: the action of the former breaks the input-state symmetry for the latter, preventing the its compression.

{\it Benefits} -- The minimum CNOT counts for a double-excitation term was 13 in~\cite{ynam_NPJ_water}. For bosonic, the count reduced to two~\cite{ynam_NPJ_water}.
If a hybrid-enabled reduction is applicable, the count becomes seven (See Fig.\ref{fig:hybrid_circuit_template}(a)). Thus, finding an optimal ordering of the excitation terms to be applied in our ansatz circuit that enable maximal compression via symmetry preservation becomes an important task.

\begin{figure}[h]
\includegraphics[scale=0.133]{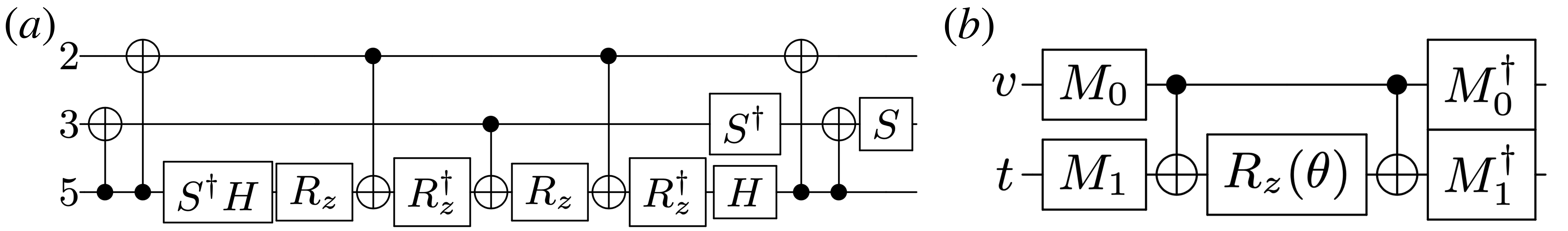}
\caption{(a) Optimized circuit for the term $U=e^{\theta_{2356} c_2^\dagger c_3^\dagger c_5 c_6 - h.c.}$.
The $R_z$ gate used in the circuit refers to $R_z(\theta_{2356}/2)$.
The qubit indices are labeled on the left of the circuit.
(b)
Template circuit for the unitary $U=e^{-i \frac{\theta}{2} {\sigma}_0^v \otimes {\sigma}_1^t \otimes \cdots}$, where $t$ denotes the target qubit location and $v$ denotes  control qubit locations. ${\sigma}_i \in \{ X, Y,Z \}$, and  $M_i \in \{ H,S^\dagger H, \mathbf{1}\}$. If ${\sigma}_i^j = X$, $M_i^j = H$. If ${\sigma}_i^j = Y$, $M_i^j = S^\dagger H$. If ${\sigma}_i^j = Z$, $M_i^j = \mathbf{1}$.
} 
\label{fig:hybrid_circuit_template}
\label{fig:pauli_circuit_template}
\end{figure}

{\it Solution} -- In general, bosonic and hybrid encoding can be considered for any pair of spin orbits. Here, we limit the consideration to the spin degree of freedom, i.e. $c^\dagger_{p} c^\dagger_{p+1}$ for odd $p$, since most significant physical excitation terms tend to preserve such symmetry. The symmetry-preserving ordering problem is now ready to be mapped to the well-known graph vertex coloring problem (GVCP). We do this in three steps: Graph construction, reduction, and coloring. 

{\it Graph construction} -- Consider a set of hybrid terms that are the  rotation associated with $h_i = c^\dagger_{p_i} c^\dagger_{p_i+1} c_{q_i} c_{r_i}$ (or also $h_i = c^\dagger_{p_i} c^\dagger_{q_i} c_{r_i} c_{r_i+1}$ if the index ordering matters as in the UCC ansatz), where $1 \leq i \leq N_h$ and $N_h$ is the number of hybrid excitation terms. The goal is to maximize the number of hybrid terms that admit compressed implementation. To achieve this, we first map the set of the hybrid terms to a directed graph $G$. In particular, the graph $G$ contains $N_h$ vertices, where each vertex corresponds to one hybrid excitation term $h_i$. We assign a directed edge from $h_j$ to $h_k$ if $h_j$ rotation breaks the parity symmetry required for $h_k$ rotation. Specifically, we use a {\it sufficient} condition 
for $h_i$ breaking the parity symmetry required for $h_j$, i.e.,
\[
\small
   B(h_i, h_j) = (q_i=p_j) \vee (r_i=p_j) \vee (q_i=p_j+1) \vee (r_i=p_j+1),
    \label{eq_hybrid_condition}
\]
where $\vee$ denotes logical or operation and the directed edge from $h_i$ to $h_j$ exists if the logical $B(h_i,h_j)$ evaluates to true. Concretely,
\[
    G = (V, E), 
    V = \{ h_i | 1 \leq i \leq N_h \},
    E = \{(h_i, h_j) | \text{ if } B(h_i, h_j)  \},
\]
where $(x,y)$ denotes a directed edge from $x$ to $y$.

{\it Graph reduction} -- 
In this step, we consider the source and sink vertices of $G$. Sink vertices $S_{sink}$ are the vertices that do not have any outgoing edge. Source vertices $S_{source}$ are the vertices that do not have any incoming edge. By design, the sink vertices of the graph $G$ do not break any parity symmetry used for hybrid-excitation compression. Therefore, all the sink vertices are implemented first. For the source vertices, the parity symmetry needed is always preserved regardless of which hybrid excitation terms were implemented beforehand. Therefore, all the source vertices are implemented last. Remove the sink and source vertices from $G$. Repeat the removal until there is no more sink or source vertices left.

{\it Graph coloring} -- We start with a reduced $G$ without any source or sink. Note, not all the hybrid excitation terms in the reduced graph may be implemented with compression, since implementing one with compression may break the parity symmetry required for another. In order to thus maximize the number of hybrid excitation terms that can be implemented with compression, we use GVCP. 

Start by constructing an undirected graph via removing the direction information in the reduced graph $G$. Label each vertex with a different color such that no two vertices sharing the same edge have the same color, while minimizing the number of colors used. In other words, for a given undirected graph $G = (V, E)$, we aim to find its chromatic number $k$ and the coloring of each vertex when only $k$ colors are used. Note every vertex with the same color can be implemented with compression. By minimizing the number of colors used, then finding the largest set of vertices with the same color, we obtain a heuristic solution to the problem of finding the maximal number of vertices that can be implemented with compression. 

Summing up, referring the maximal set of vertices that can have the same color as $S_{color}$, we convert sets $S_{sink}$, $S_{color}$, and $S_{source}$ to circuits $C_{sink}$, $C_{color}$, and $C_{source}$, respectively, using the procedure described in the next section. The resulting circuit is of the form $C = C_{source}C_{color}C_{sink}$. The hybrid terms or vertices that lie outside the three sets 
are folded into the fermionic terms.

\subsection{Advanced sorting}
\label{sec_fermion_GTSP}
Instead of Trotterizing at the excitation-operator level, implementing $e^{O_k}$ one at a time, a Pauli-string level Trotterization can be applied: Expand $O_k$ according to $\theta \sum_\kappa \otimes_{i=1}^N \sigma_{\kappa,i}$, $\sigma_{\kappa,i} \in \{I,X,Y,Z\}$, then reorder the Pauli strings across {\it different} $O_k$ terms. Finding the best ordering of Pauli strings can be viewed as a traveling salesman problem (TSP), with the distance as the gate-count savings between the consecutive Pauli strings, 
whose matrix exponentiations are to be implemented one after the other. Now, allow for each Pauli string to have its own ``target qubit,'' in contrast to the baseline. Mapping each string to a cluster, and each subvertex of the cluster being a possible target, the problem becomes determining which subvertex to visit per cluster while visiting each cluster once to make a loop. Hence generalized TSP (GTSP).

Note the exact form of the string is determined by the excitation terms considered and the choice of fermion-to-qubit transformation. Assuming a set of Pauli strings are provided, our task is to determine the implementation order of the strings based on the gate cancellation that occurs in between them. We adapt a well-known 
meta-heuristic algorithm to find a locally optimal solution for the GTSP.

The input for the problem is a set of excitation terms ($E$), either the fermionic excitation terms, or a segment of the hybrid excitation terms, i.e., source, color, or sink, and a fermion-to-qubit transformation. Using the transformation, the input excitation terms are converted to a set of Pauli string operators 
$
    E \rightarrow \{P_k| 1\leq k \leq M \},
    \label{eq_trotter_terms}
$
where $M$ is the number of the Pauli strings, $P_k = \otimes_{i=1}^N \sigma_{k;i} $ and $\sigma_{k;i} \in \{I, X, Y, Z\}$. Individually exponentiated, a conventional circuit synthesis method uses $2(w-1)$ CNOT gates for each string, where $w$ is the weight of the string defined as the number of non-$I$ matrices in a given string. In particular, one chooses a target qubit to be any one of the qubits a non-$I$ Pauli matrix acts on, then uses the circuit template shown in Fig.~\ref{fig:pauli_circuit_template}(b). 

\begin{figure*}
\centering
\includegraphics[scale=0.173]{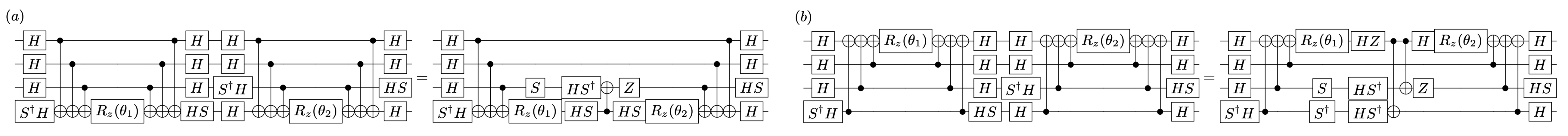}
\caption{\label{fig:pauli_circuit_simplify} Circuits for the unitary $U=e^{-i \frac{\theta_1}{2}XXXY}e^{-i \frac{\theta_2}{2}XXYX}$ with two different target qubit choices. (a) the fourth qubit. (b) the first qubit.
} 
\end{figure*}

Consider now implementing two circuit elements that correspond to two different Pauli strings in a row. 
Depending on the strings and the specific choice of the target qubit, different numbers of CNOT gates may be canceled. 
Example: Suppose we have $P_1 = XXXY$ and $P_2 = XXYX$. 
In the first scenario, we consider both strings with the common target of the fourth qubit, i.e., $t_1 = t_2 = 4$. 
At the interface of two circuits that correspond to $[P_1,t_1]$ and $[P_2,t_2]$, CNOT cancellations occur, resulting in a single CNOT (see Fig.~\ref{fig:pauli_circuit_simplify}(a)).
Let us consider the second scenario where $t_1 = t_2 = 1$. 
Then at the interface, a CNOT cancellation occurs, resulting in two CNOTs (see Fig.~\ref{fig:pauli_circuit_simplify}(b)).
Point made: The choice of target matters. Note $t_1 = t_2$ is required for CNOT cancellations.
Indeed, when given $[P_1,t_1]$ and $[P_2,t_2=t_1]$, the number of CNOT cancellations obtained at the interface may be computed as $2(w_1-1)+2(w_2-1)-\sum_i \omega_i$, where $w_k$ is the weight of the strings $P_k$, $i$ traverses every non-target qubit, and $\omega_i\in\{0,1,2\}$ is zero if any of $\sigma_{k,i}$ of $P_k$, $k\in\{1,2\}$, is identity, two if we have, on the target, any one of $(X,Y)$, $(Y,X)$, $(X,X)$, $(Y,Y)$, or $(Z,Z)$ collisions and, on the control, any one of $(X,X)$, $(Y,Y)$, or $(Z,Z)$ collisions, and one otherwise. 
See \cite{Automated,ynam_NPJ_water,hmp2} for the CNOT cancellation rules.

In order to maximize the cancellation, we map the sorting problem to GTSP. In GTSP, we are given a partition of vertices into clusters and we aim to find a minimum-length path that includes exactly one vertex from each cluster.
Note the two degrees of freedom: the ordering of the strings and the target qubit choice for each string. We assign each vertex of a weighted undirected graph $G$ with each $[P_i,t_i]$. We insert an edge if and only if $P_i \neq P_j$, and the weight of the edge is the number of CNOT reduction expected from placing the circuit blocks for $e^{i P_i \theta_i}$ and $e^{i P_j \theta_j}$ next to each other using target qubits $t_i$ and $t_j$, respectively. To sum up, the weighted undirected graph $G$ is of the form
\begin{align}
    G &= (V, E)\notag \\
    V &= \{[P_k, t_k] | 1 \leq k \leq M, 1 \leq t_k \leq N \ni \sigma_{k, t_k} \neq I  \} \notag \\
    E &= \{(v_{k_1}, v_{k_2}, d_{k_1 k_2}) | (v_{k_1}, v_{k_2}) \in V^2 \ni P_{k_1} \neq P_{k_2}  \}, \notag
    \label{eq_GTSP_graph}
\end{align}
where $(v_a, v_b, d)$ denotes an edge connecting $v_a$ and $v_b$ with weight $d$.
We now partition all the vertices $V$ in the graph $G$ into $M$ clusters i.e. $V = V_1 \cup V_2 \cup \cdots \cup V_M$. The $j$th cluster contains all the vertices with the same $P_j$ and all possible target qubit indices. 
In order to maximize the CNOT reduction, the objective is to find a maximum weight cycle containing exactly one vertex from each cluster. This problem is equivalent to solving the GTSP.

The method we use to solve GTSP
is as follows. We first multiply the weight of each edge by $-1$ so that the minimum path length corresponds in the graph $G$ corresponds to the maximum CNOT reduction. If we find the optimal path $X^*$, we can convert it to the optimal permutation (the order in which each cluster is visited) and the optimal target qubit choices (the specific vertex visited in each cluster). With the ordering and the target qubits used determined, the ansatz circuit $C$ is determined.

\subsection{Advanced fermion-to-qubit transformation}
\label{sec_search_gamma}

Consider $\Gamma {\in} {\rm GL}(N,2)$, the group of binary $N{\times}N$ invertible matrices, in contrast to being restricted to an upper-triangular one as in the baseline. For $x\in \mathbb{F}_2^N$, where $\mathbb{F}_2$ is the binary field, $\Gamma x$ is also in $\mathbb{F}_2^N$. Indeed, $\Gamma$ denotes a linear reversible circuit, a subset of Clifford~\cite{bravyi2022constant}. Since Clifford maps a pauli string to another, $U_\Gamma$, a unitary implied by $\Gamma$, maps a pauli string $P_k$ to another pauli string $U_\Gamma P_k U_\Gamma^\dagger$.

One of the commonly used fermion-to-qubit transformation is the Jordan-Wigner (JW) transformation:
$
    c_i = (\bigotimes_{j < i} Z_j) \sigma^-_i,
$
where $Z_j$ here is the Pauli-$Z$ matrix on qubit $j$. Applying the transformation to fermionic excitation operators, one obtains pauli strings. Make no further changes with $\Gamma = 1$: This is the JW transformation. In the advanced fermion-to-qubit transformation, $\Gamma \neq 1$ is used to enable the best optimization in the steps considered previously.

{\it Problem} -- The choice of $\Gamma$ depends heavily on the relatively small number of excitation terms considered, since asymptotically Bravyi-Kitaev (BK) transformation and ternary tree transformation \cite{jiang2020optimal} are already optimal. Further, the search space for $\Gamma$ is prohibitively large ($\sim 2^{N^2}$), even for a modest number of qubits $N$. 

{\it Solution} -- We exploit the topology formed by the excitation terms, i.e., we keep track of index pairs of neighboring fermion operators in the creation or annihilation part of every fermionic double excitation terms. 
This way, we can construct a graph with its vertices being the indices and its edges being the index pairs. The graph constructed, if disjoint, is divided into disjoint pieces, where each piece contains a vertex set. We use these sets as blocks in a block diagonal $\Gamma$.

To illustrate, let us consider an $N{\times}N$ $\Gamma$ that is block diagonal in blocks of size $N/2 {\times} N/2$. Given a Pauli string $P_x$ obtained from using the JW transformation, the block diagonal $\Gamma$ transforms the first half of $P_x$ separately from the latter half of $P_x$, since $P_x \mapsto U_\Gamma P_x U_\Gamma^\dagger$.
While it may not guarantee a globally optimal solution, the block diagonal $\Gamma$ provides a way to locally optimize within each block. 
Indeed, each block can be searched separately, thereby reducing the search problem size dramatically.

For each block, we use simulated annealing (SA), instead of particle swarm optimization that tends to get stuck in local minima, to find an optimal $\Gamma$. Briefly, inspired by the heating and cooling processes of a thermodynamic system, SA is a procedure to approximate the global optimum of a function, $f(x)$, over a discrete search space $x \in \mathcal{X}$. In particular, in analogy to the thermodynamic free energy, we use the Metropolis-Hasting sampling algorithm in the $\mathcal{X}$ space such that $\pi(x) \propto \exp(-f(x)/T)$, where $\pi(x)$ is the unnormalized probability distribution and $T$ is a parameter in analogy to the temperature in the thermodynamics. We can then gradually reduce the temperature $T$ to $T = 0$ such that the probability of sampling the global minimum becomes overwhelmingly large compared to other configurations. Therefore, the global minimum can be approximated. 

\comment{ }

%%%%%%%%%%%%%%%%%
\section{Circuit optimization results}
\label{sec:Results}

\begin{figure}
\centering
\includegraphics[scale=0.6]{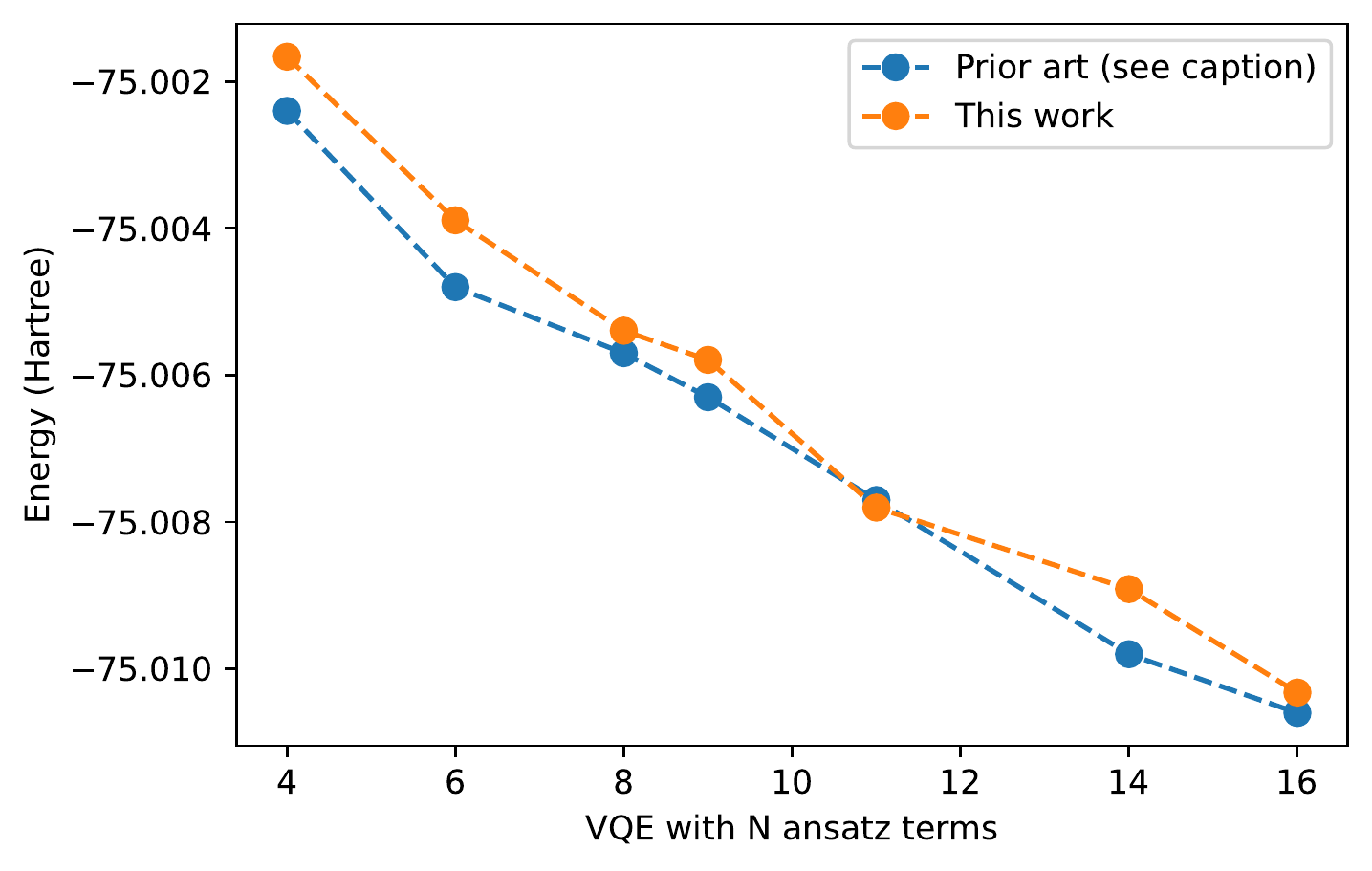}
\caption{\label{fig:convergence} Ground-state energy estimate of the water molecule as a function of the number of ansatz terms in STO-3G basis. Blue: Energies reported in \cite{hmp2}. Orange: Energies obtained via the methodology reported in this  paper. Observed is that the energies obtained between the two methods are comparable. The number of ansatz term sufficient to reach the chemical accuracy is 17 (not shown) for both methods.}
\end{figure}

In this section, we present the result for the optimized CNOT gate counts for a number of common molecules. We first generate the excitation terms using HMP2 algorithm proposed in \cite{hmp2}. We then use our advanced compilation and optimization methodology to synthesize optimal circuits. 

Before we present the optimized results, we first provide the implementation details of the optimization solvers we used in our subroutines. In the hybrid encoding subroutine, one of our tasks is to search for an optimal solution for the graph coloring problem. We use a randomized, greedy coloring algorithm to minimize the number of distinct colors. Specifically, we color the vertices in multiple different orders, where the different orders are generated at random. Each time we assign a color to a vertex, we bias our selection, choosing the same color(s) as much as possible. A new color is added, only if needed. Our randomized greedy coloring algorithm thus explores beyond a local minimum, since we randomly generate several coloring orders and return the best found solution. In the advanced sorting subroutine, a GTSP solver is required. We use the genetic algorithm (GA) to solve the GTSP problem\footnote{Well-known heuristics such as Lin-Kernighan (LK)~\cite{lin1973effective}, used to obtain a good solution to the famous TSP, can indeed be adapted to handle the GTSP~\cite{karapetyan2011lin,helsgaun2015solving}. However, the modifications needed to obtain good solutions to the GTSP are often non-trivial; For example, a 2-opt, LK solution of a GTSP, obtained by assuming a pre-determined set of vertices (one per cluster) to visit, can be one of the worst paths to visit each cluster, the moment we change the vertices to visit. We therefore leave the modified Lin-Kernighan approach as a future work and use here instead a simpler-to-implement GA.} as described in \cite{silberholz2007generalized}. 

In Table~\ref{table:cnot_count}, we show the CNOT gate counts of the optimized VQE circuits, simulating a variety of molecules. Compared therein are the CNOT gate counts obtained with different fermion-to-qubit transformations considered previously, including the JW, BK, and the generalized transformation (GT)~\cite{hmp2}, along with our methodology.

The savings obtained by our advanced compilation and optimization vary in the suite of molecules we consider, ranging from 3.56\%\footnote{The relatively small improvement ratio may be due to the inefficiency of the heuristic solvers used or the absence of a significantly better solution than known. Preliminary results (not shown) suggest the former to blame in part, with frequent local minima trapping. Another possibility may be the problem formulation itself, where the molecular orbital basis may be better chosen to reflect the inherent symmetry of the molecule.} (NH$_3$)
to 24.00\% (HF, LiH). Our methodology is capable of further optimizing the quantum circuits over the previous state of the art obtained by the GT approach \cite{hmp2}. 
These savings come with no hidden costs or accuracy loss in energy estimates. 
Indeed, we show in Fig.~\ref{fig:convergence} the convergence of the ground-state energy estimates for the water molecule, comparing the results obtained by our advanced approach to those reported in \cite{hmp2}. We confirm the energy convergence hardly changes, evidenced by the fact that both approaches still use 17 excitation terms to reach chemical accuracy.

\begin{table}[h]
\caption{Number of CNOT gates used for the VQE simulation of different molecules with different
fermion to qubit transformations} 
\centering
\begin{threeparttable}

\begin{tabular}{|c |c |c |c |c |c |c |c|} 
\hline
 Molecule& $N_e$& JW & BK & GT & Adv & Improve(\%) \\ [0.5ex] 
 \hline\hline
 HF & 3& 30& 29& 25& 19& 24.00\\
 \hline
 LiH  & 3& 30& 29& 25& 19& 24.00\\
 \hline
 BeH$_2$  & 9& 70& 71& 60& 53& 11.67\\
 \hline
 NH$_3$  & 52& 485& 607& 478& 461& 3.56\\
 \hline
 H$_2$O(4)  & 4& 42& 50& 33& 27&18.18\\
 \hline
 H$_2$O(5)  & 5& 44& 52& 35& 29&17.14\\
 \hline
 H$_2$O(6) & 6& 46& 47& 37& 31&16.21\\
 \hline
 H$_2$O(8) & 8& 68& 88& 63& 50&20.63\\
 \hline
 H$_2$O(9) & 9& 71& 89& 66& 53&19.69\\
 \hline
 H$_2$O(11) & 11& 93& 110& 87& 67&22.98\\
 \hline
 H$_2$O(12) & 12& 95& 112& 89& 70&21.34\\
 \hline
 H$_2$O(14) & 14& 114& 140& 111& 88&20.72\\
 \hline
 H$_2$O(16) & 16& 135& 166& 131& 105&19.85\\
 \hline
 H$_2$O(17) & 17& 137& 168& 133& 107&19.55\\
 \hline
\end{tabular}
\begin{tablenotes}
\item Note: $N_e$ is the number of excitation
terms considered in the UCCSD ansatz. 
JW/BK are the number of two-qubit gates with Jorden-Wigner and Bravyi-Kitaev
transformations. 
GT is the number of two-qubit gates reported in \cite{hmp2}. The ansatz terms are determined according to the HMP2 ordering with STO-3G basis set and ground state geometry, detailed in \cite{hmp2}.
In the first four rows, we report the results for the cases where the HMP2 method was used to reach chemical
accuracy. 
In the the rest of the rows, we report the results for the HMP2 progression for a water molecule. 
The number in the parentheses next to H$_2$O indicates the total number of excitation terms $N_e$ considered
for the UCCSD ansatz. 
The last result, H$_2$O(17) achieves the chemical accuracy. 
\end{tablenotes}
\end{threeparttable}
\label{table:cnot_count}
\end{table}

\comment{ }

\section{Discussion and Outlook}
\label{sec:Outlook}

While we focus on the compilation and optimization of the UCCSD ansatz, we note that our framework extends to other type of fermionic unitary operations. For example, the advanced sorting routines is immediately applicable to qubit coupled-cluster (QCC) method. Moreover, one can also extend our optimization framework to the quantum simulation of time evolution of a fermionic system. 

It is intriguing to further explore the fermion-to-qubit transformation by studying the Clifford conjugation of the Jordan-Wigner transformation, or encoding one fermionic degree of freedom with multiple qubits \cite{chen2018exact}, instead of GL($N$) we explored in this work. This may be of tremendous practical importance to future modular hardware running Hamiltonian simulations, since we may be able to use the Clifford transformation, pay upfront as a one-time cost, to ``localize'' the interactions to between ``nearby,'' connected modules only. This way, inter-modular connection resources, expected to be very expensive~\cite{Nickerson_2014,Arnold_2020,leung2019deterministic},
can be saved by a significant amount. 

\section*{ACKNOWLEDGEMENTS}
This work is supported by the ARO through the
IARPA LogiQ program; the NSF STAQ and QLCI programs; the DOE QSA program; the AFOSR MURIs
on Dissipation Engineering in Open Quantum Systems,
Quantum Measurement/ Verification, and Quantum Interactive Protocols; the ARO MURI on Modular Quantum Circuits; and the DOE HEP QuantISED Program.

\bibliographystyle{IEEEtran}
\bibliography{main_arxiv.bib}
\appendix

\subsection{Example - hybrid encoding}
\label{app:hybrid}
In this section we demonstrate our procedures to categorize hybrid terms.
Suppose we have nine hybrid double excitation fermionic terms $\{h_0,h_1,\cdots,h_8\}$
such that the directed graph, constructed according to Sec.~\ref{sec_hybrid_encoding} is that shown in Fig.~\ref{fig:example-hybrid-encoding}(a).
One example is 
\begin{align}
h_0{=}a_{9}^\dagger a_{12}^\dagger  \bm{a_3a_4},\,
&h_1{=}\bm{a_{11}^\dagger a_{12}^\dagger} a_3a_6,\,
h_2{=}a_{20}^\dagger a_{21}^\dagger\bm{a_5 a_6},\, \nonumber \\
h_3{=}a_{19}^\dagger a_{22}^\dagger \bm{a_5 a_{6}},\,
&h_4{=}a_{13}^\dagger a_{16}^\dagger\bm{a_1 a_{2}},\,
h_5{=}a_{11}^\dagger a_{14}^\dagger\bm{a_{5} a_{6}},\,
\nonumber \\
h_6{=}\bm{a_{13}^\dagger a_{14}^\dagger} {a_{5} a_{8}},\,
&h_7{=}a_{13}^\dagger a_{16}^\dagger \bm{a_{7} a_{8}},\,
h_8{=}\bm{a_{17}^\dagger a_{18}^\dagger} {a_3a_8}.\,
\nonumber
\end{align}
Notice the part in bold font can potentially be implemented with compression.
Taking `Sink' and `Source' terms out iteratively, then dropping the directedness of the edges of the graph, we obtain an undirected graph, shown in Fig.~\ref{fig:example-hybrid-encoding}(b).

Next, to detail the coloring procedure, we show two random orders in Fig.~\ref{fig:example-hybrid-encoding}(c). The two random orders are: Order 1 $h_1{-}h_5{-}h_0{-}h_6{-}h_7$ and Order 2 $h_1{-}h_7{-}h_6{-}h_5{-}h_0$.
For Order 1, we color the first term $h_1$ with blue. 
Since the next term $h_5$ is connected to $h_1$, we add a new color red.
The third term $h_0$ is connected to $h_1$ but not $h_5$. 
Since we minimize the number of colors used, we reuse the red color for $h_0$.
$h_6$ is connected with $h_5$, but not $h_1$ nor $h_0$, so we reuse the blue color. 
Lastly, $h_7$ neighbors $h_6$ only, we can reuse color red.
A similar procedure can be employed for Order 2.
Here, for $h_5$, it is connected to both $h_1$ and $h_6$, whose respective colors are blue and red. Thus, a third color cyan is added.

According to Fig.~\ref{fig:example-hybrid-encoding}(c), Order 1 contains the largest color set which is red.
As a result, we have $S_{sink}=\{h_2, h_3\}$, $S_{source}=\{h_4, h_8\}$, $S_{color}=\{h_0, h_5,h_7\}$ implemented with compression while $\{h_1, h_6\}$ are considered as regular fermionic terms.

\begin{figure}[h]
\centering
\includegraphics[scale=0.29]{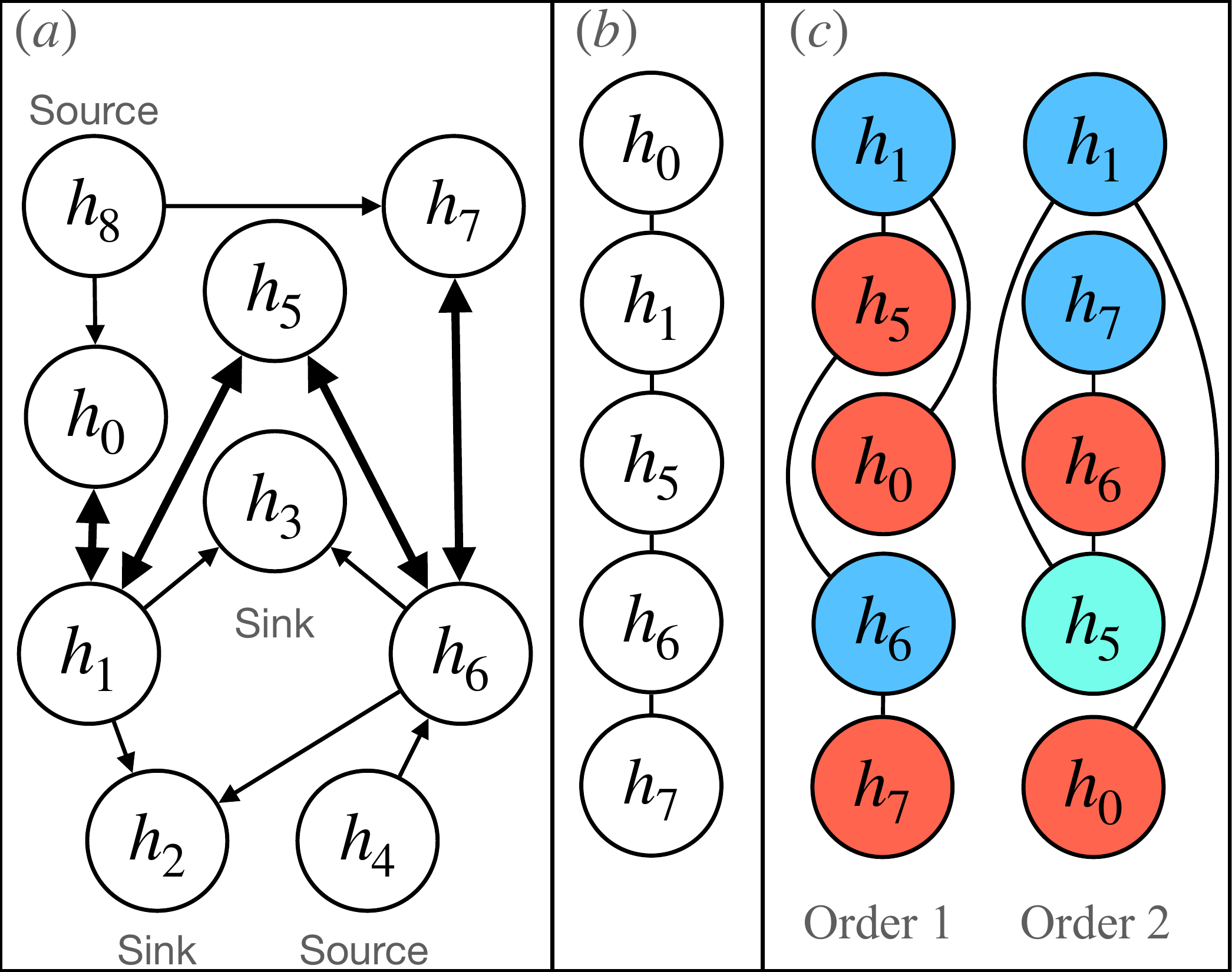}
\caption{\label{fig:example-hybrid-encoding} Example procedure to categorize hybrid terms using the graph vertex coloring problem. 
See Sec.~\ref{app:hybrid} for detailed description.
(a) {\it Graph construction} -- Directional graph for nine hybrid double excitation terms where the direction $h_i\rightarrow h_j$ implies term $h_i$ breaks the parity symmetry of $h_j$.
(b) {\it Graph reduction} -- After removing the sink and source vertices iteratively, the directedness of the edges of the remaining graph is dropped.
(c) {\it Graph coloring} -- 
The coloring order is from up to down. The lines denote the edges and the pair of vertices adjoined by an edge must be be colored differently. } 
\end{figure}

\subsection{Example - advanced sorting}
Recall in our advanced sorting, we first construct a weighted, undirected graph, given a set of Pauli strings. 
Consider an example, where $P_0{=}IIXXYXII$, $P_1{=}IIXXXYII$, and $P_2{=}XXIIIIXY$. The valid target qubits are then
$\{3,4,5,6\},\{3,4,5,6\}$, and $\{1,2,7,8\}$, respectively. The graph constructed can then be visualized as in Fig.~\ref{fig:example-advanced-sorting} (edges not shown for visual clarity).
A valid path is a series of connected edges, where each cluster (Pauli string) is visited exactly once.   
The weight $d$ of an edge $([P_j,t],[P_k,t'],d)$ is computed as the negative of the number of CNOT gates cancelled when implementing $e^{- i \theta P_j/2}$ and $e^{- i \theta' P_k/2}$ one after the other with target qubits $t$ and $t'$, respectively, according to the method detailed in Sec.~\ref{sec_fermion_GTSP}.
In this example, among all $4^3$ possible paths, we aim to find a path that minimizes the path weight, defined as the sum of the weights of the edges in the path.
Our sorting problem is thus fully mapped to the GTSP problem.

\begin{figure}
\centering
\includegraphics[scale=0.3]{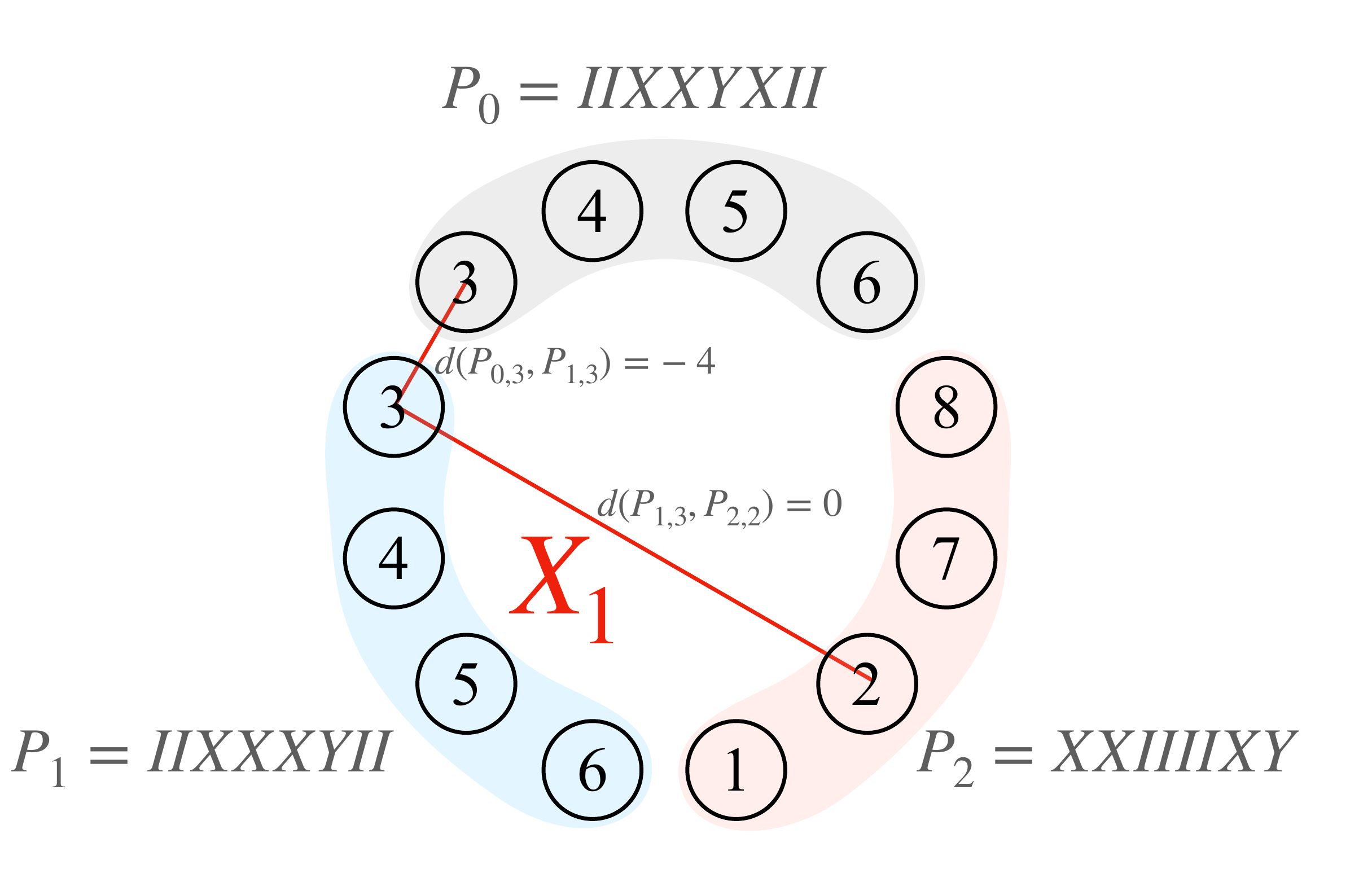}
\caption{ \label{fig:example-advanced-sorting} GTSP graph constructed for our example advanced sorting problem.
Each of the circled numbers inside the same color block (cluster) represents a valid target qubit index of Pauli string $P_j$.
A valid path traverses each cluster once, visiting one circled number per cluster, such as in $X_1:([P_{0},3], [P_{1},3], [P_{2},2])$.
The path weight of $X_1$ is the sum of the weights of the edges in the path, i.e., $([P_{0},3],[P_{1},3],-4)$ and $([P_{1},3],[P_{2},2],0)$. The path weight is thus $-4$.
}
\end{figure}

\subsection{Example - Optimization for $\Gamma$}
We show in this section an example where (i) we reduce the search space of $\Gamma$ based on the excitation term topology and (ii) the effect of the choice of a different $\Gamma$ on the pauli strings whose matrix exponentiations are to be implemented.
To illustrate (i), consider two excitation terms $a_9^\dagger a_8^\dagger a_3a_1$ and $ a_6^\dagger a_5^\dagger a_2a_1$. The creation part has connected clusters $\{8,9\}$ and $\{5,6\}$, while the annihilation part has a connected cluster $\{1,2,3\}$.
Therefore our $\Gamma$ matrix candidate may consist of a $3\times 3$ block for indices  $\{1,2,3\}$ and two $2\times 2$ blocks for indices  $\{8,9\}$ and $\{5,6\}$ respectively.
To illustrate (ii), consider an example Pauli string $P=XXIIXY$, say, obtained as a part of the JW transformation of some fermionic excitation term. 
For a block diagonal $\Gamma$ candidate, consider
\begin{align}
    \Gamma = \left[
    \begin{array}{cc|cc|cc}
    1   &0  &0   &0   &0  &0 \\
    1   &1  &0   &0   &0   &0 \\ \hline
     0   &0   & 1 &0   &0  &0 \\
     0   &0     & 0  & 1  &0  &0 \\ \hline
     0   &0     & 0  & 0   &1  &0 \\
      0  &0     & 0  & 0   &1  &1 
    \end{array}\right],
\end{align}
where we explore non-identity upper left and lower right $2{\times}2$ blocks in the hopes to change the input pauli string to a form more amenable to better circuit optimization by the rest of our circuit optimization procedures. 
The particular $\Gamma$ shown above corresponds to applying CNOTs on the first two and the last two qubits, which transforms $P$ as 
$(CNOT\otimes II\otimes CNOT )\times (XXIIXY) \times ( CNOT \otimes II \otimes CNOT) = XIIIYZ$, which is a different and shorter Pauli string.
The search space for the candidate $\Gamma$ here is indeed a much smaller one to explore than that for the full $6{\times}6$ matrix in our current example.
Our block-diagonal-based strategy can be employed to an arbitrary-sized generalized transformation we consider, i.e., a linear reversible circuit, which can be implemented using \cite{patel2008optimal}.

\end{document}